\documentclass[
	a4paper, % Paper size, use either a4paper or letterpaper
	10pt, % Default font size, can also use 11pt or 12pt, although this is not recommended
	unnumberedsections, % Comment to enable section numbering
	twoside, % Two side traditional mode where headers and footers change between odd and even pages, comment this option to make them fixed
]{LTJournalArticle}

\runninghead{} % A shortened article title to appear in the running head, leave this command empty for no running head

\footertext{} % Text to appear in the footer, leave this command empty for no footer text

\title{Virtual memory for real-time systems using hPMP} % Article title, use manual lines breaks (\\) to beautify the layout

\author{%
	Konrad Walluszik\textsuperscript{1}, Daniel Auge\textsuperscript{1}, Gerhard Wirrer\textsuperscript{1}, Holm Rauchfuss\textsuperscript{1} and Thomas Röcker\textsuperscript{1}
}

\date{\footnotesize\textsuperscript{\textbf{1}}Infineon Technologies AG, 85579 Neubiberg}

\renewcommand{\maketitlehookd}{%
	\begin{abstract}
		\noindent To satisfy automotive safety and security requirements, memory protection mechanisms are an essential component of automotive microcontrollers. In today's available systems, either a fully physical address-based protection is implemented utilizing a memory protection unit, or a memory management unit takes care of memory protection while also mapping virtual addresses to physical addresses. The possibility to develop software using a large virtual address space, which is agnostic to the underlying physical address space, allows for easier software development and integration, especially in the context of virtualization. In this work, we showcase an extension to the current RISC-V SPMP proposal that enables address redirection for selected address regions, while maintaining the fully deterministic behavior of a memory protection unit.
	\end{abstract}
}

\usepackage{xcolor}
\begin{document}

\maketitle % Output the title section

%----------------------------------------------------------------------------------------
%	ARTICLE CONTENTS
%----------------------------------------------------------------------------------------
\section{Introduction}
In automotive compute systems, the aspect of memory protection plays a distinctly important role for satisfaction of safety- and security requirements. From a type-classification perspective, we distinguish mechanisms employing physical memory only vs. virtual memory based systems. The latter applying any kind of translation, i.e. effective addresses are ‘mapped’ to physical addresses based on a certain ruleset. Applying a selective set based strategy, large physical memories can be addressed by virtual address ranges.
In virtual memory based systems, typically translation is performed page based, i.e. blocks of predefined size (and located at certain physical addresses) get a virtual address assigned, and can consequently form contiguous (or non-contiguous) address maps though the individual blocks might physically scattered \cite{b0}. Such an approach implies the need for a look-up mechanism, which is invoked upon every access; consequently, latency is induced to the system. In order to mitigate this effect, caching strategies can be applied which avoids potentially costly multi-stage lookups (also referred as table lookups). While the caching increases performance of lookup \textit{on average}, it becomes an additional burden when trying to analyze worst-case timings/boundaries due to the induced dependency on execution history.

Considering growing SW-footprints and agile divide-and-conquer approaches in large-scale developments, agnosticism to physical addressing allows easier SW-integration along the whole lifecycle (deploy-run-invalidate-update-run). Importance of this mechanism is pronounced in particular when deploying multiple MCU images to a single physical controller: these 'virtual MCUs' (vMCU) are then mapped to virtual harts/machines (VM), which are managed by a hypervisor (HV). The VMs target logical independence, i.e. physical address agnosticism (deploy), freedom from interference (run), secure VM atomic image replacement (invalidate-update).

\section{Requirements for realtime virtualization}
In order to leverage the benefits of virtual memory for automotive applications with realtime requirements, a solution is mandated which provides an address-translation feature minimizing additional complexity introduced to analysis of time boundaries. (Note: The authors acknowledge that usage of virtualization of every manner adds complexity over purely physical solutions, yet at the benefit of reduced hardware cost.) Furthermore, the feature should be transparent to applications/setups for which virtual addressing is not required, i.e. change of the programming model of the standard memory protection unit shall be avoided. Finally, impact to memory access timing needs to be avoided, especially when considering systems using fast local memories.

\begin{figure}
    \centering
    \includegraphics[width=0.4\linewidth]{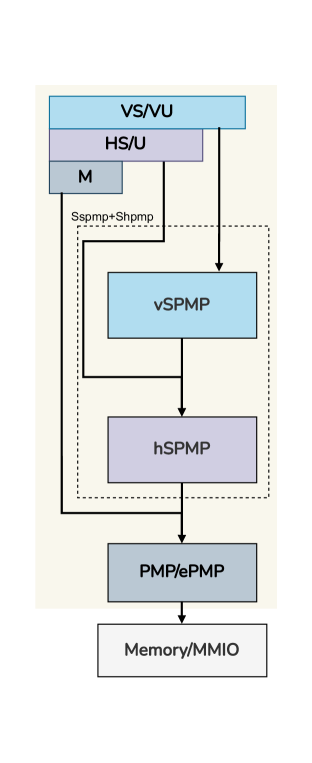}
    \caption{Two level physical memory protection \cite{b3}}
    \label{fig:2level_pmp}
\end{figure}

RISC-V specifies virtualization support in the privileged architecture specification by introducing the hypervisor (H) extension \cite{b1}: The supervisor privilege level is 'split' into hypervisor-extended supervisor (HS) and virtual supervisor mode (VS). On top of the virtual supervisor mode there is the virtual user mode (VU-mode) introduced. To address the realtime need for separation of VMs, a two-level configuration of RISC-V SPMP \cite{b2} is required, as proposed in work \cite{b3}: The first stage, named vSPMP, is in control by the guest operating systems running in VS-Mode of a VM. The second stage, called hSPMP or hPMP, is controlled by the hypervisor HS privilege level, enforcing isolation between VMs (Figure \ref{fig:2level_pmp} illustrates the approach). 

For our extension, we assume the following model: Address translation is only performed by hPMP, i.e. both guest- and user-code operate on guest physical addresses employing a dedicated memory protection unit. In order to integrate address translation, our proposed extension requires a modification of the baseline hypervisor, which manages the translation rule-set.

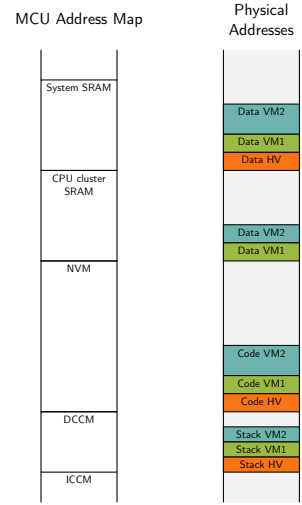
\begin{figure}[htbp]
    \centering
    \begin{tikzpicture}[scale=0.4, every node/.style={scale=0.4, /utils/exec={\sffamily}} ]

    % ================ Definitions ================
    \def\width{2.5}
    \def\xOffset{6}

    \definecolor{cVM2}{rgb}{0.4235, 0.7059, 0.6784}
    \definecolor{cVM1}{rgb}{0.6078, 0.7294, 0.2627}
    \definecolor{cHV}{rgb}{0.9765, 0.4549, 0.0784}
    \definecolor{cBG}{rgb}{0.9490, 0.9490, 0.9490}

    % left, bottom, width, height, text
    \def\region(#1,#2,#3,#4,#5){
        \draw[] (#1,#2) rectangle (#1+#3,#2+#4);
        \node[anchor=north, align=center] at (#1+#3/2,#2+#4) {#5};
    }

    % left, bottom, width, height, text, fill color
    \def\regionFilled(#1,#2,#3,#4,#5,#6){
        \filldraw[fill=#6] (#1,#2) rectangle (#1+#3,#2+#4);
        \node[anchor=north] at (#1+#3/2,#2+#4) {#5};
    }

    % ================ Drawing ================
    % Left
    \node[align=center, style={scale=1.5}] at (\width/2,15) {MCU Address Map};
    \draw[] (0,13) -- ++(0,1);
    \draw[] (\width,13) -- ++(0,1);
    \region(0,10,\width,3,System SRAM)
    \region(0,7,\width,3,CPU cluster\\SRAM)
    \region(0,2,\width,5,NVM)
    \region(0,0,\width,2,DCCM)
    \node[anchor=north, align=center] at (\width/2,0) {ICCM};
    \draw[] (0,0) -- ++(0,-1);
    \draw[] (\width,0) -- ++(0,-1);

    % HV
    \node[align=center, style={scale=1.5}] at (\xOffset+\width/2,15) {Physical\\Addresses};
    \fill [cBG] (\xOffset,-1) rectangle ++(\width,15);
    \draw[] (\xOffset,-1) -- ++(0,15);
    \draw[] (\xOffset+\width,-1) -- ++(0,15);
    
    \regionFilled(\xOffset,11.2,\width,1.0,Data VM2,cVM2);
    \regionFilled(\xOffset,10.6,\width,0.6,Data VM1,cVM1);
    \regionFilled(\xOffset,10,\width,0.6,Data HV,cHV);

    \regionFilled(\xOffset,7.6,\width,0.6,Data VM2,cVM2);
    \regionFilled(\xOffset,7,\width,0.6,Data VM1,cVM1);

    \regionFilled(\xOffset,3.2,\width,1.0,Code VM2,cVM2);
    \regionFilled(\xOffset,2.6,\width,0.6,Code VM1,cVM1);
    \regionFilled(\xOffset,2,\width,0.6,Code HV,cHV);

    \regionFilled(\xOffset,1,\width,0.5,Stack VM2,cVM2);
    \regionFilled(\xOffset,0.5,\width,0.5,Stack VM1,cVM1);
    \regionFilled(\xOffset,0,\width,0.5,Stack HV,cHV);

\end{tikzpicture}
    \caption{Physical memory protection from the perspective of a hypervisor}
    \label{fig:addressmap}
\end{figure}

To illustrate the motivation behind the proposed extension, Figure \ref{fig:addressmap} (left) presents an example of an address map for an embedded microcontroller, showing regions of closely coupled memory (CCM) designated for storing instructions (I) and data (D). For storing program code and static data, a nonvolatile memory (NVM) region is considered. Furthermore, the MCU can include different, often non-contiguously addressed SRAM regions which can be utilized to store data during execution. A dedicated segment (e.g. placed at the top of the address space) is utilized for peripheral access. The right part of Figure \ref{fig:addressmap} illustrates a minimalistic example of two VMs being managed by a HV. Considered are dedicated sections in DCCM used as stack, code regions in the NVM range and two non-shared global data regions scattered across available SRAM. 

Based on the exemplary defined memory map \ref{tab:hpmp_config}, our target configuration of hPMP is employing pairs of \textit{pmpaddr}-registers, which have defined matching-mode OFF and TOR in \textit{pmpcfg}, respectively. The \textit{pmpaddr}-pairs define start and end address of a protection region (columns 'hPMP start/end address') with permissions being defined by \textit{pmpcfg} holding matching mode TOR. During reconfiguration, we consider respective ranges to be disabled by using the \textit{pmpswitch}-register. We note that hPMP-implementations might be restricted to the OFF-TOR case (effectively only supporting A=0 and A=0,1 for even- and odd-numbered cfgs, respectively). It should be noted that in case of applying statically defined protection-sets to larger, contiguously addressable memory other models might be more efficient in terms of register-/range usage.

The targeted permission model reads as follows: To all regions for the hypervisor (HV) VM-access is disallowed, protecting the HV from unintended modification and/or elevation of privilege. The VM-regions are configured with RW/RX for data and code related regions, respectively. Considering the whitelisting-logic applied in the unified-model of hPMP, this requires rules with S=0 being used. Via \textit{hpmpswitch} register, the HV will disable regions of all other VMs before scheduling next distinct VM (its regions are activated accordingly). The HV-execution itself is protected (e.g. to mitigate effects resulting from random hardware-faults) by dedicated ranges with RW/RX, yet using rules of type S=1 (activated in hpmpswitch permanently).

In the following paragraph a switching scenario will be described.
\begin{table*}[ht]
    \centering
    \resizebox{\textwidth}{!}{%
        \begin{tabular}{|c|c|c|c|c|c|c|}
\hline
Region & Name & Size [KB] & hPMP start address 'A' & hPMP cfg A & hPMP end address 'B' & hPMP cfg B \\
\hline
0 & Stack HV & 2 & hpmpaddr0 = 0x2000\_0000 & OFF & hpmpaddr1 = 0x2000\_07FF & S TOR RW \\
\hline
1 & Stack VM1 & 4 & hpmpaddr2 = 0x2000\_0800 & OFF & hpmpaddr3 = 0x2000\_17FF & - TOR RW \\
\hline
2 & Stack VM2 & 4 & hpmpaddr4 = 0x2000\_1800 & OFF & hpmpaddr5 = 0x2000\_27FF & - TOR RW \\
\hline
3 & Code HV & 256 & hpmpaddr6 = 0x8000\_0000 & OFF & hpmpaddr7 = 0x8003\_FFFF & S TOR RX \\
\hline
4 & Code VM1 & 512 & hpmpaddr8 = 0x8004\_0000 & OFF & hpmpaddr9 = 0x800B\_FFFF & - TOR RX \\
\hline
5 & Code VM2 & 512 & hpmpaddr10 = 0x800C\_0000 & OFF & hpmpaddr11 = 0x8013\_FFFF & - TOR RX \\
\hline
6 & Data VM1 & 128 & hpmpaddr12 = 0x9000\_0000 & OFF & hpmpaddr13 = 0x9001\_FFFF & - TOR RW \\
\hline
7 & Data VM2 & 128 & hpmpaddr14 = 0x9002\_0000 & OFF & hpmpaddr15 = 0x9003\_FFFF & - TOR RW \\
\hline
8 & Data HV & 96 & hpmpaddr16 = 0x9080\_0000 & OFF & hpmpaddr17 = 0x9081\_7FFF & S TOR RW \\
\hline
9 & Data VM1 & 256 & hpmpaddr18 = 0x9081\_8000 & OFF & hpmpaddr19 = 0x9085\_7FFF & - TOR RW \\
\hline
10 & Data VM2 & 256 & hpmpaddr20 = 0x9085\_8000 & OFF & hpmpaddr21 = 0x9089\_7FFF & - TOR RW \\
\hline
\end{tabular}
    }
    \vspace{12pt}
    \caption{hPMP configuration for OFF-TOR couples based on an exemplary address map}
    \label{tab:hpmp_config}
\end{table*}
For switching from one VM under execution to another VM, a call to HV needs to be triggered (e.g. by an interrupt generated by a timer). When the HV executes, it has to perform a set of macro-tasks:
\begin{itemize}
    \item Saving the state of the current VM (includes CPU registers, program counter, CPU flags, etc.)
    \item Loading the next VM state according to the VM scheduler (includes loading CPU registers, reprogramming hPMP, etc.)
    \item Execution of the next VM
\end{itemize}
To maintain a consistent CPU state it is essential that reconfigurations will be done in an atomic way without interruptions. Therefore, the hypervisor software will execute the second step, loading the next VM state, as part of a critical section where special measures are taken. For critical sections, the hypervisor is typically temporally disabling interrupts to ensure a critical sequence not being preempted. Furthermore, memory barrier or fence instructions are used to ensure the needed instructions are executed in the write order and all instructions of the critical section are done before the hypervisor moves on to the next step (Note: Explicit serialization is required, when indirect CSRs are used). Updating the hPMP configuration during VM switch can be more or less complex, depending on the number of implemented hPMP entries, the number of VMs running on the CPU and also the number and placement of memory regions to be separated for each VM. All the former aspects lead to the need for a larger number of PMP ranges/entries. In cases where the number of needed hPMP entries for all regions to be separated is smaller or equal the number of implemented hPMP entries, the update can be performed using the \textit{hpmpswitch} registers as indicated above.

\section{RISC-V CPU extension}
In our proposal, we extend the HV context by introducing additional HV-CSRs named \textit{hpmpoffsetx} (where x=0-63), encoding most significant bits in 34-bit address space of RV32. When executing in V=1, each \textit{hpmpoffsetx} is used to derive physical addresses from \textit{hpmpaddrx}, considering a hit in a respective entry of hPMP. Contrary when V=0, the registers have no effect. In this sense, Guest-OS and -applications operate on guest-physical addresses, while HV is employing physical addressing.

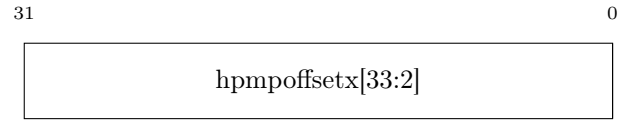
\begin{figure}[htbp]
\begin{centering}
  \begin{tikzpicture}
    % Overall width of the register (95% of column width)
    \def\registerWidth{0.95\columnwidth}
    
    % Bit field sizes (in percentage of register width)
    \def\paddressWidth{1.0} % 32 bits out of 32 (full width)

    % Register outline
    \draw (0,0) rectangle (\registerWidth,1);

    % Separation lines for bit fields
    \draw (\registerWidth,0) -- (\registerWidth,1); % Right side
    
    % Labels for bit fields
    \node[anchor=center] at (0.5*\paddressWidth*\registerWidth,0.5) {hpmpoffsetx[33:2]};

    % Bit number labels
    \node[anchor=south] at (\registerWidth,1.2) {\scriptsize 0};
    \node[anchor=south] at (0,1.2) {\scriptsize 31};

  \end{tikzpicture}
\end{centering}
\caption{hPMP hpmpoffsetx register layout, RV32}
\label{fig1:hpmpoffset}
\end{figure}

In order to keep the mechanism lean, we assume hPMP-implementations to use of OFF-TOR strategy as described above. Then even-numbered offsets can be hardwired '0', while odd-numbered \textit{hpmpoffsetx} applies translation to addresses stored in registers number \textit{x} and \textit{x-1} of \textit{hpmpaddr} (unified model). Consequently, VMs can be 'moved' within virtual address space, while resizing requires change of respective \textit{hpmpaddr}. The layout of the introduced register is shown in Figure \ref{fig1:hpmpoffset}. For a guest physical address (GPA) which has a hit within a defined hPMP region \( k \), the physical address (PA) can be calculated using the following formula:

\[
PA = GPA + \text{hpmpoffset}{\langle x \rangle}
\]

where \( x \) is calculated as:

\[
x = 2k + 1
\]
It is important to mention that even in unified model there is no impact to the HV, as offsets are not applied for HV regions itself. 
In the following, we discuss two scenarios that make use of the hPMP offsets.

\subsection{Partial Update}
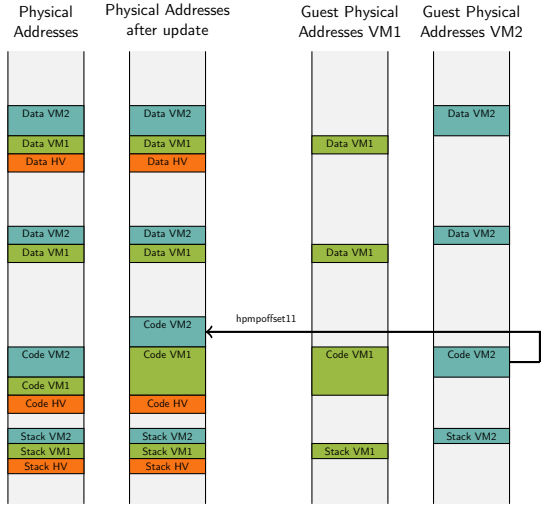
\begin{figure}[htbp]
    \centering
    \begin{tikzpicture}[scale=0.4, every node/.style={scale=0.4, /utils/exec={\sffamily}} ]

    % ================ Definitions ================
    \def\width{2.5}
    \def\xOffset{0}
    \def\xOffsetResized{4}
    \def\xOffsetVM{10}
    \def\xOffsetVMtwo{14}

    \definecolor{cVM2}{rgb}{0.4235, 0.7059, 0.6784}
    \definecolor{cVM1}{rgb}{0.6078, 0.7294, 0.2627}
    \definecolor{cHV}{rgb}{0.9765, 0.4549, 0.0784}
    \definecolor{cBG}{rgb}{0.9490, 0.9490, 0.9490}

    % left, bottom, width, height, text
    \def\region(#1,#2,#3,#4,#5){
        \draw[] (#1,#2) rectangle (#1+#3,#2+#4);
        \node[anchor=north, align=center] at (#1+#3/2,#2+#4) {#5};
    }

    % left, bottom, width, height, text, fill color
    \def\regionFilled(#1,#2,#3,#4,#5,#6){
        \filldraw[fill=#6] (#1,#2) rectangle (#1+#3,#2+#4);
        \node[anchor=north] at (#1+#3/2,#2+#4) {#5};
    }

    % ================ Drawing ================

    % HV
    \node[align=center, style={scale=1.5}] at (\xOffset+\width/2,15) {Physical\\Addresses};
    \fill [cBG] (\xOffset,-1) rectangle ++(\width,15);
    \draw[] (\xOffset,-1) -- ++(0,15);
    \draw[] (\xOffset+\width,-1) -- ++(0,15);
    
    \regionFilled(\xOffset,11.2,\width,1.0,Data VM2,cVM2);
    \regionFilled(\xOffset,10.6,\width,0.6,Data VM1,cVM1);
    \regionFilled(\xOffset,10,\width,0.6,Data HV,cHV);

    \regionFilled(\xOffset,7.6,\width,0.6,Data VM2,cVM2);
    \regionFilled(\xOffset,7,\width,0.6,Data VM1,cVM1);

    \regionFilled(\xOffset,3.2,\width,1.0,Code VM2,cVM2);
    \regionFilled(\xOffset,2.6,\width,0.6,Code VM1,cVM1);
    \regionFilled(\xOffset,2,\width,0.6,Code HV,cHV);

    \regionFilled(\xOffset,1,\width,0.5,Stack VM2,cVM2);
    \regionFilled(\xOffset,0.5,\width,0.5,Stack VM1,cVM1);
    \regionFilled(\xOffset,0,\width,0.5,Stack HV,cHV);

    % HV resized
    \node[align=center, style={scale=1.5}] at (\xOffsetResized+\width/2,15) {Physical Addresses\\after update};
    \fill [cBG] (\xOffsetResized,-1) rectangle ++(\width,15);
    \draw[] (\xOffsetResized,-1) -- ++(0,15);
    \draw[] (\xOffsetResized+\width,-1) -- ++(0,15);

        \regionFilled(\xOffsetResized,11.2,\width,1.0,Data VM2,cVM2);
    \regionFilled(\xOffsetResized,10.6,\width,0.6,Data VM1,cVM1);
    \regionFilled(\xOffsetResized,10,\width,0.6,Data HV,cHV);

    \regionFilled(\xOffsetResized,7.6,\width,0.6,Data VM2,cVM2);
    \regionFilled(\xOffsetResized,7,\width,0.6,Data VM1,cVM1);

    \regionFilled(\xOffsetResized,4.2,\width,1.0,Code VM2,cVM2);
    \regionFilled(\xOffsetResized,2.6,\width,1.6,Code VM1,cVM1);
    \regionFilled(\xOffsetResized,2,\width,0.6,Code HV,cHV);

    \regionFilled(\xOffsetResized,1,\width,0.5,Stack VM2,cVM2);
    \regionFilled(\xOffsetResized,0.5,\width,0.5,Stack VM1,cVM1);
    \regionFilled(\xOffsetResized,0,\width,0.5,Stack HV,cHV);

    % VM1 resized
    \node[align=center, style={scale=1.5}] at (\xOffsetVM+\width/2,15) {Guest Physical\\Addresses VM1};
    \fill [cBG] (\xOffsetVM,-1) rectangle ++(\width,15);
    \draw[] (\xOffsetVM,-1) -- ++(0,15);
    \draw[] (\xOffsetVM+\width,-1) -- ++(0,15);

    \regionFilled(\xOffsetVM,10.6,\width,0.6,Data VM1,cVM1);
    \regionFilled(\xOffsetVM,7,\width,0.6,Data VM1,cVM1);
    \regionFilled(\xOffsetVM,2.6,\width,1.6,Code VM1,cVM1);
    \regionFilled(\xOffsetVM,0.5,\width,0.5,Stack VM1,cVM1);

    %VM2 resized
    \node[align=center, style={scale=1.5}] at (\xOffsetVMtwo+\width/2,15) {Guest Physical\\Addresses VM2};
    \fill [cBG] (\xOffsetVMtwo,-1) rectangle ++(\width,15);
    \draw[] (\xOffsetVMtwo,-1) -- ++(0,15);
    \draw[] (\xOffsetVMtwo+\width,-1) -- ++(0,15);

    \regionFilled(\xOffsetVMtwo,11.2,\width,1.0,Data VM2,cVM2);
    \regionFilled(\xOffsetVMtwo,7.6,\width,0.6,Data VM2,cVM2);
    \regionFilled(\xOffsetVMtwo,3.2,\width,1.0,Code VM2,cVM2);
    \regionFilled(\xOffsetVMtwo,1,\width,0.5,Stack VM2,cVM2);

    \draw[<-, thick] (\xOffsetResized+\width, 4.7) -| (\xOffsetVMtwo+\width+1, 3.7);
    \draw[-, >=stealth, thick] (\xOffsetVMtwo+\width+1, 3.7) -- (\xOffsetVMtwo+\width, 3.7);
    \node[anchor=north, align=center] at (\xOffsetResized+\width+2, 4.7+0.7) {hpmpoffset11};
    
\end{tikzpicture}
    \caption{Address map with resizing feature}
    \label{fig:addressmap_remapped}
\end{figure}

In the first scenario, the code section of VM1 increases its memory footprint from 512KB to 768KB (see first vs. second column in Figure \ref{fig:addressmap_remapped}). This can be the result of an update or feature extension for the application running on this virtual machine. As a consequence it is required to move the image of VM2 to have non-overlapping address regions. Instead of rebuilding/-linking the application of VM2 (resulting in impact to adjancent VM-images and consequently effort for ECU revalidation), we employ the proposed concept of the hPMP to redirect all addresses of VM2 by a fixed offset (see arrow in Figure \ref{fig:addressmap_remapped}).

Table \ref{tab:config_update} gives an overview of the affected hPMP registers and their new values based on the discussed scenario.

\begin{table}[h!]
    \centering
    \begin{tabular}{|c|c|}
        \hline
        \textbf{Affected register} & \textbf{New value} \\
        \hline
        hpmpaddr9 & 0x800F\_FFFF \\
        \hline
        hpmpoffset11 & 0x8\_0000\\
        \hline
    \end{tabular}
    \caption{Table showing affected hPMP registers and their new values}
    \label{tab:config_update}
\end{table}

\subsection{Generic Images}
\begin{figure}[ht]
    \centering
    \begin{tikzpicture}[scale=0.4, every node/.style={scale=0.4, /utils/exec={\sffamily}} ]

    % ================ Definitions ================
    \def\width{2.5}
    \def\xOffset{4}
    \def\xOffsetVM{10}
    \def\xOffsetVMtwo{14}

    \definecolor{cVM2}{rgb}{0.4235, 0.7059, 0.6784}
    \definecolor{cVM1}{rgb}{0.6078, 0.7294, 0.2627}
    \definecolor{cHV}{rgb}{0.9765, 0.4549, 0.0784}
    \definecolor{cBG}{rgb}{0.9490, 0.9490, 0.9490}

    % left, bottom, width, height, text
    \def\region(#1,#2,#3,#4,#5){
        \draw[] (#1,#2) rectangle (#1+#3,#2+#4);
        \node[anchor=north, align=center] at (#1+#3/2,#2+#4) {#5};
    }

    % left, bottom, width, height, text, fill color
    \def\regionFilled(#1,#2,#3,#4,#5,#6){
        \filldraw[fill=#6] (#1,#2) rectangle (#1+#3,#2+#4);
        \node[anchor=north] at (#1+#3/2,#2+#4) {#5};
    }

    % ================ Drawing ================
    % Left
    % HV
    \node[align=center, style={scale=1.5}] at (\xOffset+\width/2,15) {Physical\\Addresses};
    \fill [cBG] (\xOffset,-1) rectangle ++(\width,15);
    \draw[] (\xOffset,-1) -- ++(0,15);
    \draw[] (\xOffset+\width,-1) -- ++(0,15);
    
    \regionFilled(\xOffset,11.2,\width,1.0,Data VM2,cVM2);
    \regionFilled(\xOffset,10.6,\width,0.6,Data VM1,cVM1);
    \regionFilled(\xOffset,10,\width,0.6,Data HV,cHV);

    \regionFilled(\xOffset,7.6,\width,0.6,Data VM2,cVM2);
    \regionFilled(\xOffset,7,\width,0.6,Data VM1,cVM1);

    \regionFilled(\xOffset,3.2,\width,0.6,Code VM2,cVM2);
    \regionFilled(\xOffset,2.6,\width,0.6,Code VM1,cVM1);
    \regionFilled(\xOffset,2,\width,0.6,Code HV,cHV);

    \regionFilled(\xOffset,1,\width,0.5,Stack VM2,cVM2);
    \regionFilled(\xOffset,0.5,\width,0.5,Stack VM1,cVM1);
    \regionFilled(\xOffset,0,\width,0.5,Stack HV,cHV);

    % VM1
    \node[align=center, style={scale=1.5}] at (\xOffsetVM+\width/2,8) {Guest Physical\\Addresses VM1};
    \fill [cBG] (\xOffsetVM,-1) rectangle ++(\width,8);
    \draw[] (\xOffsetVM,-1) -- ++(0,8);
    \draw[] (\xOffsetVM+\width,-1) -- ++(0,8);
    \regionFilled(\xOffsetVM,4.6,\width,0.6,Data VM1,cVM1);
    \regionFilled(\xOffsetVM,4,\width,0.6,Data VM1,cVM1);
    \regionFilled(\xOffsetVM,1,\width,0.6,Code VM1,cVM1);
    \regionFilled(\xOffsetVM,0,\width,0.5,Stack VM1,cVM1);

    % VM2
    \node[align=center, style={scale=1.5}] at (\xOffsetVMtwo+\width/2,8) {Guest Physical\\Addresses VM2};
    \fill [cBG] (\xOffsetVMtwo,-1) rectangle ++(\width,8);
    \draw[] (\xOffsetVMtwo,-1) -- ++(0,8);
    \draw[] (\xOffsetVMtwo+\width,-1) -- ++(0,8);
    \regionFilled(\xOffsetVMtwo,4.6,\width,1.0,Data VM2,cVM2);
    \regionFilled(\xOffsetVMtwo,4,\width,0.6,Data VM2,cVM2);
    \regionFilled(\xOffsetVMtwo,1,\width,0.6,Code VM2,cVM2);
    \regionFilled(\xOffsetVMtwo,0,\width,0.5,Stack VM2,cVM2);

    % Arrows avoiding boxes
    \draw[->, thick] (\xOffsetVM+\width, 1.3) -- ++(0.5, 0) -- ++(0, 1.3) -- ++(-6.5, 0);
    \draw[->, thick] (\xOffsetVMtwo+\width, 1.5) -- ++(0.5, 0) -- ++(0, 1.7) -- ++(-10.5, 0);

\end{tikzpicture}
    \caption{Address map with two virtual machines using generic images}
    \label{fig:memory_map_zero}
\end{figure}
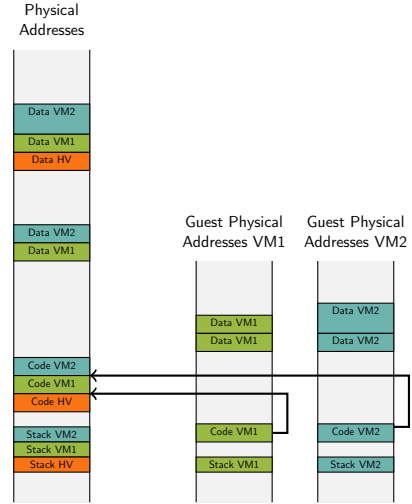

The second use case considers virtual machines that were built with the feature of address relocation in mind. The virtual machines are created in a generic way without assumptions of the memory layout in the final deployment. This will be particularly useful when addressing individualized cars with different combinations of applications for a given ECU in each car instance. With the final integration, the guest physical addresses are relocated to the actual physical locations by the hPMP offsets. In this way, generic virtual machines can be deployed on any target ECU with compatible hardware without the need to rebuild with the final addresses (see Figure \ref{fig:memory_map_zero}). For programming the hPMP this results into having multiple entries with the same guest physical address programmed into hPMP address A/B, while only in combination with the hPMP offset B values, the defined regions end up in different physical address locations. The different region and offset definitions are swapped by the hypervisor during a switch between the virtual machines. Alternatively, the \textit{hpmpswitch} register can be used to quickly enable or disable the affected hPMP entries.

\section{Conclusion}
With the introduction of the Supervisor Physical Memory Protection (SPMP) and the proposals for a two-level Physical Memory Protection approach, RISC-V is defining fundamental isolation and protection concepts for virtualized automotive systems. This work highlights the limitations of classical virtual memory using a Memory Management Unit (MMU) approach within the context of real-time and deterministic systems. Through the demonstrated hpmp extension, we introduce an address redirection feature that enables the use of virtual memory while preserving the deterministic behavior of today's memory protection units. We currently focus on studying the behavior in corner cases (e.g. overlapping regions, other matching modes, etc.), feasibility (area, power, timing), and integration to hypervisor.

%----------------------------------------------------------------------------------------
%	 REFERENCES
%----------------------------------------------------------------------------------------

\end{document}